\documentclass[aps,prl,showpacs,amsmath,amssymb,amsfonts,superscriptaddress,lengthcheck,twocolumn,longbibliography]{revtex4-1}
\usepackage{dcolumn}    
\usepackage{bm} 
\usepackage{graphicx}
\usepackage{amsmath}    
\usepackage{latexsym}
\usepackage{amsfonts}   
\usepackage{amssymb}
\usepackage{array}      
\usepackage{epsfig}
\usepackage{txfonts}
\usepackage{xcolor}
\usepackage[colorlinks=true,linkcolor=blue,urlcolor=blue,citecolor=blue,pdfusetitle]{hyperref}
\usepackage{hyperref}
\usepackage{ulem}

\newcommand{\blah}{blah\\blah\\blah\\blah\\blah.}
\newcommand{\steve}[1]{{\color{black}#1}}

\newcommand{\bra}[1]{\left\langle #1\right|}
\newcommand{\ket}[1]{\left|#1\right\rangle}
\newcommand{\braket}[2]{\left\langle #1|#2\right\rangle}

\newcommand{\la}{\left\langle}
\newcommand{\ra}{\right\rangle}

\newcommand{\bla}{bla\\bla\\bla\\bla\\bla}

\newcommand{\mc}[1]{\mathcal{#1}}

\newcommand{\mf}[1]{\mathfrak{#1}}
\newcommand{\mrm}[1]{\mathrm{#1}}

\begin{document}
\title{Orthogonality Catastrophe as a Consequence of the Quantum Speed Limit}
       
\author{Thom\'as Fogarty}
\affiliation{Quantum Systems Unit, Okinawa Institute of Science and Technology Graduate University, Onna, Okinawa 904-0495, Japan}
\author{Sebastian Deffner}
\affiliation{Department of Physics, University of Maryland, Baltimore County, Baltimore, MD, 21250, USA}
\author{Thomas Busch}
\affiliation{Quantum Systems Unit, Okinawa Institute of Science and Technology Graduate University, Onna, Okinawa 904-0495, Japan}
\author{Steve Campbell}
\affiliation{School of Physics, University College Dublin, Belfield Dublin 4, Ireland}

\begin{abstract}
A  remarkable feature of quantum many-body systems is the orthogonality catastrophe which describes their extensively growing sensitivity to local perturbations and plays an important role in condensed matter physics. Here we show that the dynamics of the orthogonality catastrophe can be fully characterized by the quantum speed limit and, more specifically, that any quenched quantum many-body system whose variance in ground state energy scales with the system size exhibits the orthogonality catastrophe. Our rigorous findings are demonstrated by two paradigmatic classes of many-body systems -- the trapped Fermi gas and the long-range interacting Lipkin-Meshkov-Glick spin model.
\end{abstract}
\date{\today}
\maketitle

\paragraph{Introduction.}

Numerous many-body systems exhibit properties and phases that cannot be explained in exclusively classical terms. Famous examples include Bose-Einstein condensation \cite{Cornell2002,Ketterle2002}, topological states \cite{Hasan2010,Bansil2016}, non-classical dispersion relations \cite{Peres2010,Armitage2018}, and many-body localization \cite{Abanin2019}, to name just a few. Whereas static properties are often understood in great detail, understanding dynamical properties can be significantly more involved. Nevertheless, it is the dynamical properties that are particularly interesting for quantum technological applications, as exemplified by quantum thermodynamic devices~\cite{SteveAndDeffner2019book}.

Mathematically, the issue is due to the fact that to describe the dynamics of many-body systems the time-dependent Schr\"odinger equation has to be solved for an immense number of microscopic variables -- which is practically unfeasible. One way forward is then to obtain qualitative insights from fundamental statements of quantum physics, which, in part, is why the study of the quantum speed limit (QSL) has spurred an area of research in its own right~\cite{DeffnerReview}. The QSL is a careful formulation of Heisenberg's uncertainty relation for energy and time \cite{Heisenberg1927} and bounds the minimal time, referred to as the QSL time $\tau_\text{QSL}$, that a quantum system needs to evolve between distinct states \cite{Mandelstam,Bhattacharyya1983,Uffink1993,margolus98,MacconePRA2003}. Originally formulated for undriven Schr\"odinger dynamics  \cite{Mandelstam,Uffink1993,margolus98,MacconePRA2003}, the QSL has been generalized to controlled \cite{Pfeifer1993,Poggi2013,Deffner2013,Hegerfeldt2013,Barnes2013a, Lychkovskiy2, PolkovnikovPRX} and open systems~\cite{delcampo13,Taddei2013,DeffnerPRL2013,deffner14,Cimmarusti2015, LidarPRL2015, Pires2016,Deffner2017NJP,BinderPRL2018}.

Interestingly, in its original inception the QSL was formulated to bound the minimal time for the evolution between two \textit{orthogonal} states~\cite{Mandelstam,margolus98}. It is therefore interesting to consider its relation to the discovery by Anderson~\cite{Anderson}  that a local perturbation on a gas of $N$ fermions causes a change in the quantum many-body states that is strongly dependent on $N$. In particular, in the limit of large $N$ the local perturbation forces the system to assume an orthogonal state -- an effect known as orthogonality catastrophe (OC). The OC has been analyzed in many different scenarios \steve{including quantum spin models~\cite{ZinardiPRE, Schiro1}, trapped gases~\cite{GooldPRA2011, SindonaPRL2013, CampbellPRA2014, Schiro2, CoscoArXiv, BuschPRL2019}, impurity models~\cite{vonDelftPRB}, it has also been explored in thermal states~\cite{DemlerPRX}}, and in understanding the breakdown in quantum adiabaticity~\cite{Lychkovskiy1}. However, to date a clear connection between the dynamics as characterized by the QSL and the orthogonality catastrophe has not been made.

In this work we aim at closing this gap in the fundamental understanding of the dynamics of quantum many-body systems. Typically the OC is characterized by the dynamical overlap $\chi(t)$, which is closely related to the Loschmidt echo~\cite{ZanardiPRL,Goussev2008}, and is defined as the inner product of the state in absence and presence of the perturbation. We show that the QSL, i.e. the maximal rate with which any quantum many-body system can evolve, is also governed by $\chi(t)$. With this fundamental relation at hand, we then conclude that the OC appears in any quantum many-body system in which the variance of the energy scales with the number of particles $N^\alpha$, where $\alpha$ is an exponent determined by the specific system properties. This conclusion is then explored and demonstrated for two important many body systems: the trapped Fermi gas and the isotropic Lipkin-Meshkov-Glick (LMG) model~\cite{LMG}, which is a paradigmatic example of strongly interacting systems~\cite{HaiTaoPRA2007,LucyEPJD2016,Vidal0,Vidal1,Vidal2,BrazilLMG,CanevaPRB,CampbellPRL2015,CampbellPRB2016}.

\paragraph{Anderson's orthogonality catastrophe.}
In his original formulation, Anderson considered the effect a local perturbation has on a gas of $N$ spinless fermions~\cite{Anderson} and showed that the overlap between the perturbed and unperturbed many-body states, written as
\begin{equation}
\chi=\braket{\Psi(x_1,x_2,\dots,x_N)}{\Phi(x_1,x_2,\dots,x_N)}\,,
\end{equation}
scales as $\chi\propto N^{-\alpha/2}$, where $\alpha$ is related to the perturbation strength. As a consequence, even a small perturbation causes two many-body states to become orthogonal as $N$ grows. Although Anderson's treatment focused on stationary states, dynamical orthogonality after sudden quenches can be similarly observed and is described by a dynamical overlap
\begin{equation}
	\label{EchoEq}
	\chi(t)= \bra{\Psi} e^{i\mathcal{H}_f t}e^{-i\mathcal{H}_i t}\ket{\Psi},
\end{equation}
with the initial state $\Psi$ being an eigenstate of the Hamiltonian $\mathcal{H}_i$, while $\mathcal{H}_f$ is the perturbed Hamiltonian~\footnote{For brevity, in the formulae, we work in units such that $\hbar\!\!=\!\!1$.}. This is related to the survival probability or time-dependent fidelity, $\mc{F}(t)\!\!=\!\!\vert \chi(t) \vert^2$, which is an important quantifier of out-of-equilibrium dynamics~\cite{Adolfo1, Adolfo2, Adolfo3, Jafari1, Jafari2, Jafari3, Jafari4}. Indeed one can find footprints of the dynamical OC in the spectral function, $S(\omega)\!\!=\!\!2\, \mf{Re}\left( \int_{-\infty}^{\infty}dt\,\chi(t)e^{i\omega t}\right)$, which is broadened by the OC and possesses a power law tail~\cite{Nozieres}. While the OC is well-known in condensed matter physics, theoretical studies have proposed using cold atomic systems to observe and study it, due to the ability to create clean many-body states with separately controllable impurity atoms~\cite{GooldPRA2011,DemlerPRX}. Recent experiments have been able to measure the survival probability and spectral function of a Fermi gas of $^6$Li after an interaction quench with $^{40}$K impurities by using a Ramsey atom-interferometric technique heralding the OC~\cite{CetinaPRL2015, CetinaScience}. 

\paragraph{``Catastrophic" quantum speed limit.}
To establish a relation between the OC and the QSL we start by inspecting the dynamical overlap, $\chi(t)$. Since $\ket{\Psi}$ is an eigenstate of the unperturbed Hamiltonian, $\mc{H}_i\ket{\Psi}\!\!=\!\!\mc{E}_i\ket{\Psi}$, we can write
\begin{equation}
 \chi(t)= \bra{\Psi} e^{i\mathcal{H}_f t}\ket{\Psi} e^{-i\mathcal{E}_i t}\,.
\end{equation}
This allows us to introduce the Bures angle between the two states $\ket{\psi_0}=\ket{\Psi}$ and $\ket{\psi_t}=e^{i\mathcal{H}_f t}\ket{\Psi}$
\begin{equation}
\mc{L}(t)\equiv \arccos{|\chi(t)|}= \arccos{|\langle \psi_0\vert \psi_t \rangle |}\;,
\end{equation}
which is only implicitly dependent on the unperturbed Hamilonian $\mc{H}_i$. At any time $\tau$, the Bures angle has an upper bound given by  \cite{Wootters1981,Taddei2013}
\begin{equation}
\label{eq:L}
\mc{L}(\tau)\leq \frac{1}{2}\int_0^\tau dt \sqrt{\mc{I}}\,,
\end{equation}
where $\mc{I}$ is the quantum Fisher information \steve{with respect to time}. For pure states and Hamiltonian dynamics it can be computed explicitly as~\cite{Boixo2007}
\begin{equation}
\mc{I}=4\left(\la \mc{H}_f^2\ra-\la \mc{H}_f\ra^2\right)=4 \Delta\mc{H}_f^2\,,
\end{equation}
and one can therefore immediately see that the dynamics, when described by the dynamical overlap, is fully characterized by the variance of the perturbed Hamiltonian, $\mc{H}_f$. Introducing now the well known connection between the QSL and the quantum Fisher information, $v_\mrm{QSL}\!\!\equiv\!\!\sqrt{\mc{I}}/2$ ~\cite{Taddei2013,Campbell2017,DeffnerReview,Safranek2018,Deffner2017NJP},
one can see that the QSL can be written as $v_\mrm{QSL}= \Delta\mc{H}_f$.  Re-substituting this into Eq.~\eqref{eq:L}, and noting that $\Delta\mc{H}_f$ is time-independent, then gives a direct connection between the QSL time and the dynamical overlap as
\begin{equation}
\label{eq:tQSL}
\tau \geq \tau_\mrm{QSL}=\frac{\arccos{|\chi(\tau)|}}{\Delta\mc{H}_f}\,.
\end{equation}
The maximal rate of quantum evolution, $v_\mrm{QSL}$, is therefore determined by the energy variance of the perturbed Hamiltonian which is a function of the number of particles $N$. As a consequence we see that $\tau_\mrm{QSL}\!\to\!0$ when $\Delta\mc{H}_f$ scales extensively with $N$, \steve{which means that the time a large system needs to evolve between two orthogonal states vanishes. We then see that the OC is a consequence of the quantum speed limit: an extensive post-quench Hamiltonian variance drives the many-body system to evolve significantly faster, and correspondingly the time to reach any orthogonal state vanishes.}

\paragraph{Orthogonality catastrophe and other QSLs.}
It is worth noting that $\chi(t)$ as given in Eq.~\eqref{EchoEq} is closely related to the thermodynamic work, $W$, performed in perturbing the many-body system. Thus far, by virtue of the sudden quench approximation, it is clear that for $t\!\geq\!0^+$ the Hamiltonian is time independent and the ensuing dynamics unitary. As such it is easy to convince oneself that the Mandelstam-Tamm bound, virtually in its original form, presents a natural choice for exploring the OC. However, this picture does not explicitly account for the switching-on of the interaction/perturbation, which necessarily requires some work to be performed~\cite{CoscoArXiv, MarchNJP2016, CampbellPRB2016}. We can explore this connection in a concrete manner by exploiting the fact that QSL times can be derived for any given distinguishability metric~\cite{Deffner2017NJP}. Choosing Eq.~\eqref{EchoEq} as our figure of merit, we can derive an alternative expression for $\tau_\mrm{QSL}$ that carries additional physical significance in terms of the work done in quenching the system~\cite{Funo2017,supplementarymaterial} and find
\begin{equation}
\label{fogartyQSL}
\tau_\mrm{W} = \frac{\hbar\left(1 - \vert\chi_\tau\vert \right)}{\vert \langle W \rangle \vert}\;,
\end{equation}
where $\langle W \rangle=\partial_t \chi(t) \vert_{t=0}$ is the average work performed due to the quench~\cite{Silva2008,Talkner2016, supplementarymaterial} \steve{and exhibits similar scaling to $\tau_\mrm{QSL}$. It is worth noting that Eqs.~\eqref{eq:tQSL} and \eqref{fogartyQSL} also demonstrate that the formalism of the QSL provides a useful framework to explore fundamental properties of any given dynamics. As the QSL is inherently dependent on which distinguishability metric is employed, closely related bounds could be derived that account for other features of the system, such as the coherence (see e.g. Ref.~\cite{DeffnerReview} for an overview). By choosing the survival probability we have established a strict relationship between the emergence of the OC and the dynamics and thermodynamics of the quench process.}

\begin{figure}[t]
 \hskip0.05\columnwidth { \bf (a)} \hskip0.43\columnwidth {\bf (b)}\\
\includegraphics[width=0.49\columnwidth]{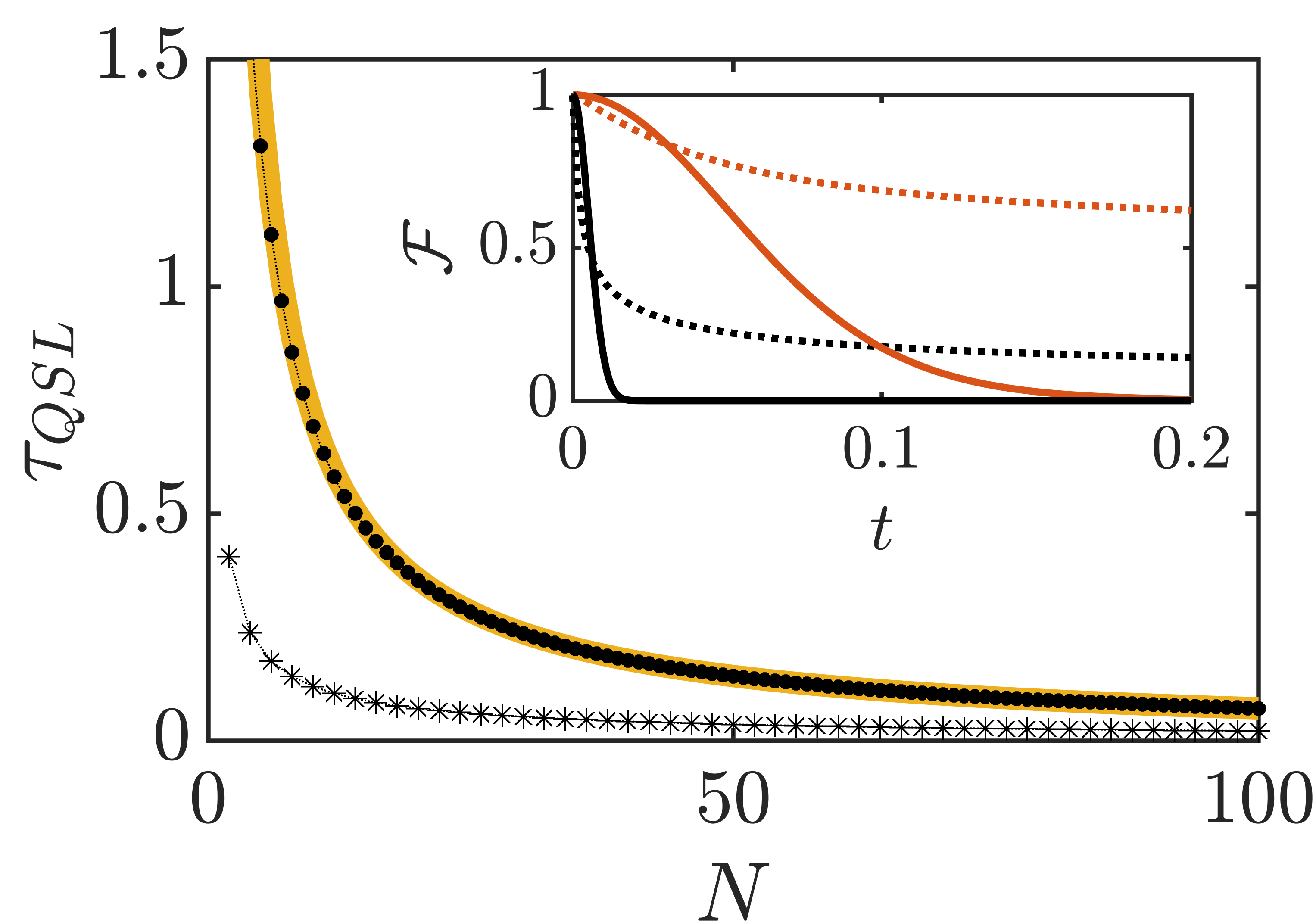}~\includegraphics[width=0.48\columnwidth]{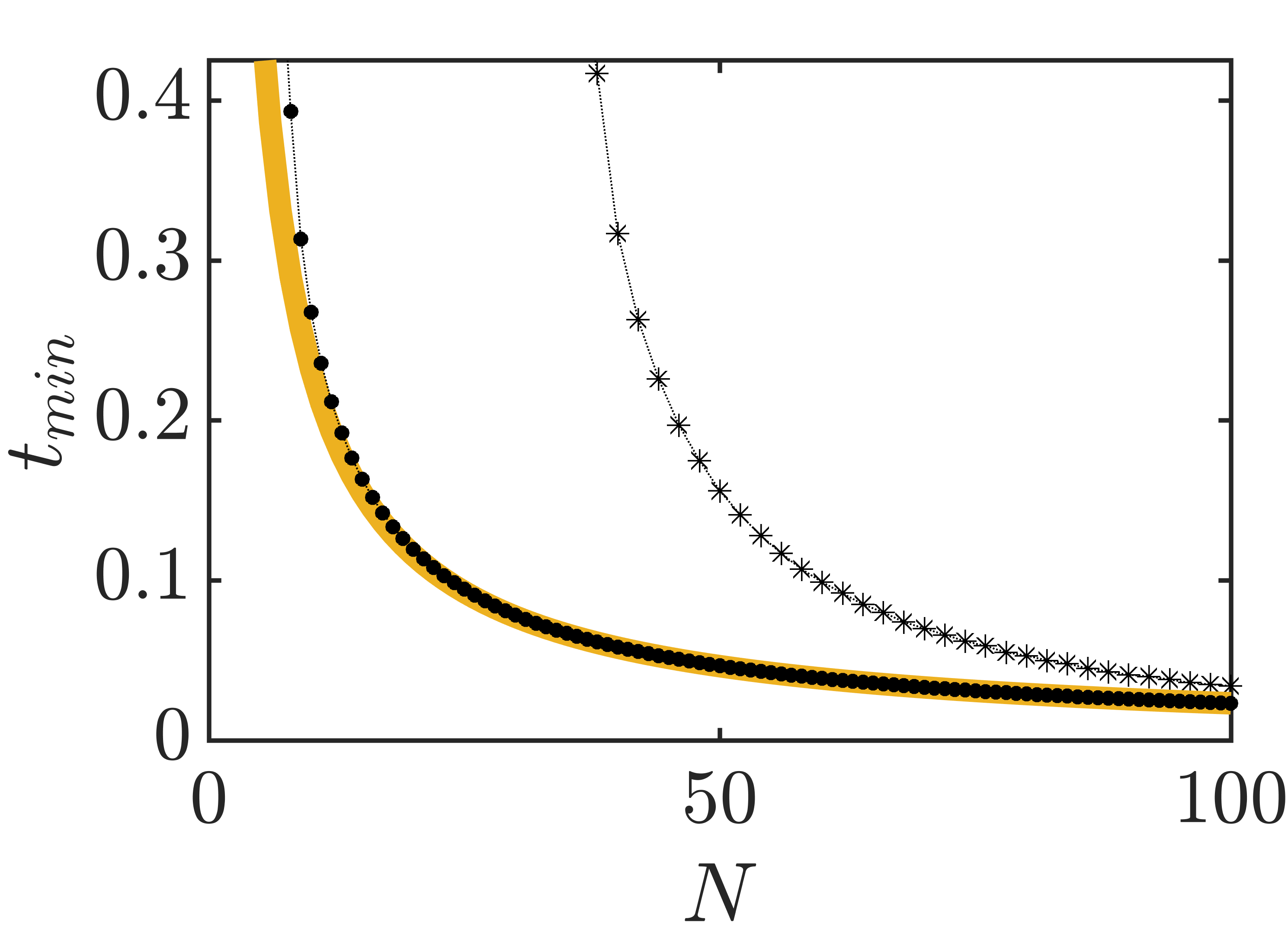}
\caption{(a) Dots: QSL time, Eq.~\eqref{eq:tQSL}, as a function of particle number for a trap-quench of strength $\eta\!\!=\!\!1.5$. The yellow line uses the approximate expression for $\Delta H$ in the large $N$ limit. Stars: QSL time for an impurity quench of strength $\kappa\!\!=\!\!0.5$. Inset: Survival probability vs. time for $N\!\!=\!\!10$ (red lines) and $N\!\!=\!\!100$ (black lines) for a trap quench (solid lines) and  an impurity quench (dotted lines). (b) Minimum time to reach $\mathcal{F}(t)\!\!=\!\!10^{-2}$ for the trap quench (dots) with the approximation in Eq.~\eqref{eq:tmin} (yellow line), and minimum time to reach $\mathcal{F}(t)\!\!=\!\!0.25$ for the impurity quench (stars).}
\label{fig:QSLw}
\end{figure}

\paragraph{Trapped Fermi Gas.}
As a first example, we now explore the above connection in a harmonically trapped Fermi gas, which is close to Anderson's original setting~\cite{Anderson}. The $N$-body wavefunction can be constructed through the Slater determinant of the respective single particle eigenstates
\begin{equation}
\Psi(x_1,x_2,\dots,x_N)=\frac{1}{\sqrt{N!}}\det^{N}_{n,j=1}[\psi_{n}(x_{j})]\;,
\end{equation}
which are in turn defined before and after a sudden quench by $\mathcal{H}_i \psi_n\!\!=\!\!E_n \psi_n$ and $\mathcal{H}_f \phi_n\!\!=\!\!E'_n \phi_n$, respectively. The survival probability of the many-body state is then 
\begin{align}
\mathcal{F}(t)=&\vert \bra{\Psi} e^{i\mathcal{H}_f t}e^{-i\mathcal{H}_i t}\ket{\Psi}\vert^2\\
		     =&\vert \det(\mathcal{A}(t)) \vert^2,
\label{eq:Ft_fermi}
\end{align}
where the elements of the matrix $\mathcal{A}$ are the overlaps of the single particle states $\psi_k(x,0)$ and $\psi_l(x,t)$ as~\cite{Anderson}
\begin{align}
\mathcal{A}_{k,l}(t)=&\int_{-\infty}^{\infty} \psi_k(x,0) \psi^*_l(x,t) dx\\
                              =&\sum_{m=1}^{\infty} \langle \psi_k \vert \phi_m\rangle \langle \psi_l \vert \phi_m\rangle e^{-i \left( E'_m-E_k \right) t}\;.
\end{align}
This significantly simplifies the calculation of $\chi(t)$ and allows one to consider large systems. Indeed, for a sudden quench in the trapping frequency  $\omega_1\!\to\!\omega_2$  such that $\eta\!=\!\omega_2/\omega_1>1$, the single particle overlaps are known analytically~\cite{Naqvi, Adolfo3}. The static \steve{(i.e. overlap with the ground state)} and dynamical survival probabilities can be calculated as
\begin{align}
\mathcal{F}=&|\langle \Psi | \Phi \rangle|^2=\left( \frac{2 \sqrt{\eta}}{\eta+1} \right)^{N^2}\\
\mathcal{F}(t)=&\left( \frac{2 \eta}{\sqrt{4 \eta^2 \cos^2(t)+(\eta^2+1)^2 \sin^2(t)}} \right)^{N^2}\;.
\label{eq:timedepfid}
\end{align}
One immediately sees that both decay with the exponent $N^2$ and depend on the strength of the quench, $\eta$. For larger systems the survival probability decays faster (see inset of Fig.~\ref{fig:QSLw}(a)) which is the manifestation of the OC.

To determine the QSL time, Eq.~\eqref{eq:tQSL}, we require $\Delta H$ for the Fermi gas which is given by
\begin{equation}
\Delta H=\frac{\eta^2-1}{2\sqrt{2} N}\sum_{n=1}^{N}\sqrt{n^2-n+1}\!\approx\!\ \frac{N}{4\sqrt{2}}\left[\eta^2-1\right]\;,
\label{eq:varN}
\end{equation}
where the approximate expression is valid for large particle numbers $N$. The QSL time therefore exhibits an extensive behavior with the system size (see Fig.~\ref{fig:QSLw}(a)) which is qualitatively similar to the survival probability.  \steve{Similarly, the average work is given by $\langle W \rangle=\frac{N}{4}\left[\eta^2-1\right]$ and exhibits  scaling comparable to $\tau_\mrm{QSL}$ \cite{MarchNJP2016}. To formally relate the QSL time and the survival probability} we calculate the minimum time for the latter to reach a specific value, i.e. $\mathcal{F}(t_{min})=\vartheta$, and find from Eq.~\eqref{eq:timedepfid} 
\begin{equation}
t_{min}=\frac{1}{\pi}\sec^{-1}\left[ \frac{\eta^2-1}{\sqrt{1+\eta^4+\eta^2 (2-4\vartheta^{-2/N^2})}} \right]\;,
\label{eq:tminExact}
\end{equation}
which for large $N$ reduces to
\begin{equation}
t_{min}\approx\frac{2\eta}{\pi N}\frac{\sqrt{\log(\vartheta^{-2})}}{\eta^2-1}\;.
\label{eq:tmin}
\end{equation}
Therefore this minimum time can be related through the energy variance in Eq.~\eqref{eq:varN} to the speed limit as
\begin{equation}
t_{min}\sim \tau_{QSL}\frac{\eta}{\pi^2}\sqrt{-\log(\vartheta)}\;,
\label{eq:approx_tmin}
\end{equation}
which shows that the QSL bounds the minimum time to reach $\mathcal{F}(\tau_\text{QSL})=e^{-\pi^4/\eta^2}$, as shown in Fig.~\ref{fig:QSLw}(b). In fact, for sudden quenches it is not surprising that the appearance of dynamically orthogonality depends on the QSL time, as the variance of the non-equilibrium excitations and the evolution of the survival probability are described by the same distribution of single particle probabilities.

We can also consider the setting first proposed by Anderson, quenching the interaction with an impurity embedded in the Fermi sea which leads to a power-law decay of the survival probability~\cite{Anderson,GooldPRA2011,DemlerPRX}. Describing the interaction with the impurity as a delta-function with a height $N \kappa$ the single particle Hamiltonian can be written as 
\begin{equation}
   H=-\frac{\hbar^2}{2m}\nabla^2+\frac{1}{2}m\omega^2 x^2+ \Theta(t)\; N \kappa\; \delta(x)\;,
\label{eq:delta}
\end{equation}
where $\Theta(t)$ is the Heaviside step function that suddenly switches on the interaction for $t\!>\!0$. Similarly to the trap quench, the QSL time and $t_{min}$ exhibit an extensive dependence on $N$ (see Fig.~\ref{fig:QSLw}), reaffirming the previous analysis.

\paragraph{Orthogonality catastrophe in interacting systems.}
We next consider a more complex setting where an impurity is immersed in an interacting bath. In particular we choose the model of a single spin interacting with a critical isotropic Lipkin-Meshkov-Glick (LMG) environment \cite{HaiTaoPRA2007, LucyEPJD2016}. The total Hamiltonian is given by $H\!\!=\!\! H_\text{LMG} + H_\text{int}$ with
\begin{equation}
\begin{split}
H_\text{LMG} &= -\frac{\lambda}{N} \sum_{i<j}^N \left( \sigma_x^i\sigma_x^j + \sigma_y^i\sigma_y^j \right) - \sum_{i=1}^N \sigma_z^i, \\
 H_\text{int} &=  \frac{\gamma}{N} \sum_{i}^N \left(  \sigma_x^i\sigma_x^s +  \sigma_y^i\sigma_y^s  \right) - \sigma_z^s.
\end{split}
\end{equation}
Here $H_\text{int}$ accounts for the impurity-bath interaction term with strength $\gamma$ and the free Hamiltonian of the impurity system, $s$. The LMG model is an example of a critical spin system that exhibits a quantum phase transition at $\lambda\!\!=\!\!1$~\cite{LMG, Vidal0, Vidal1, Vidal2, BrazilLMG, CanevaPRB, CampbellPRL2015, CampbellPRB2016}. It is convenient to work in the angular momentum basis in terms of collective spin operators, $S_\alpha \!=\! \sum_i^N \sigma_\alpha^i$. In this picture the Hamiltonian becomes
\begin{equation}
\begin{split}
H &= -\frac{\lambda}{N} \left( S_+S_- + S_-S_+ -N \openone_N \right)\\
&\quad - 2 S_z -2\frac{\gamma}{N}(s_+ S_- + s_- S_+) - 2 s_z,
\end{split}
\end{equation}
where we have also used the spin operators for the impurity. In line with the original framework of Anderson where the impurity corresponded to a small perturbation, and following the previous analysis, we will fix $\gamma\! \!=\!\! \lambda \sqrt{N}$ such that the impurity interacts comparatively weakly with the bath.

 \begin{figure}[t]
 {\bf (a)} \hskip0.45\columnwidth {\bf (b)}\\
\includegraphics[width=0.48\columnwidth]{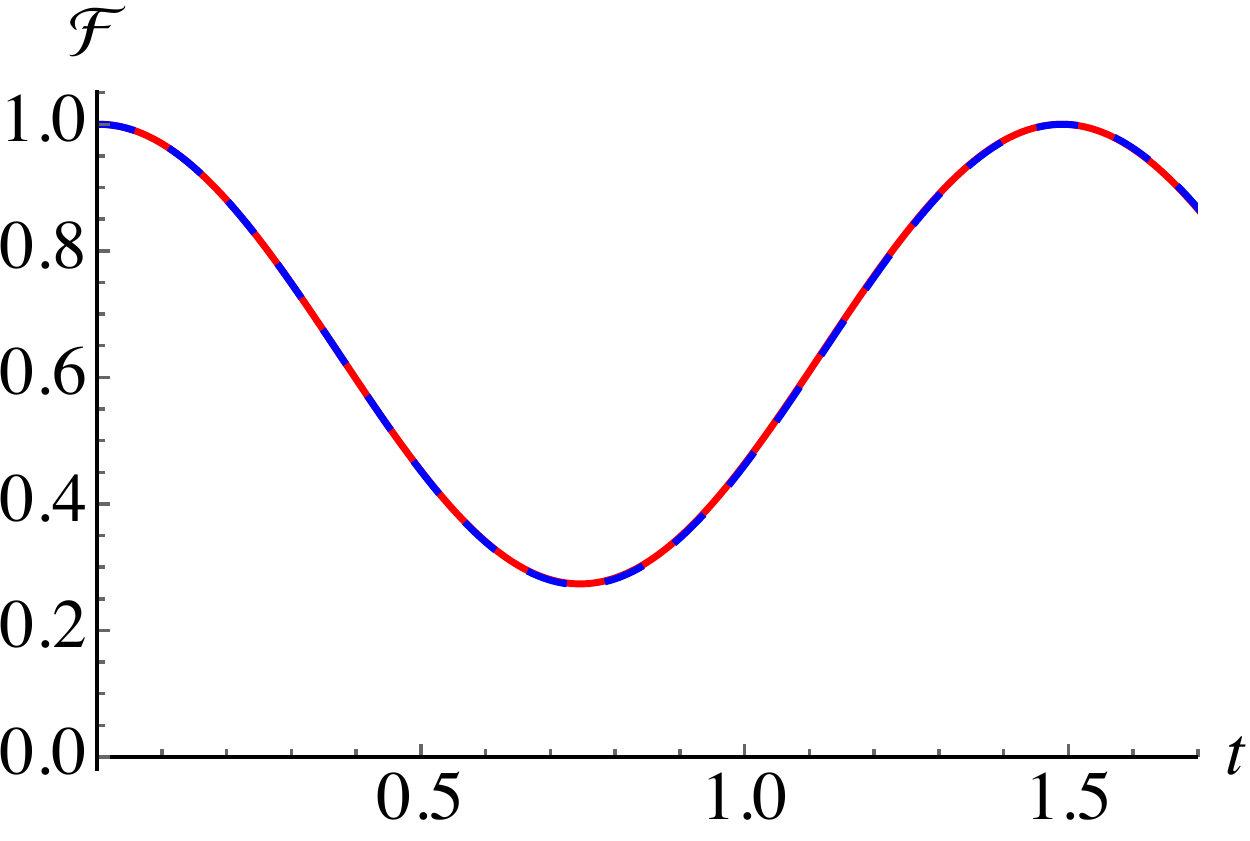}~\includegraphics[width=0.48\columnwidth]{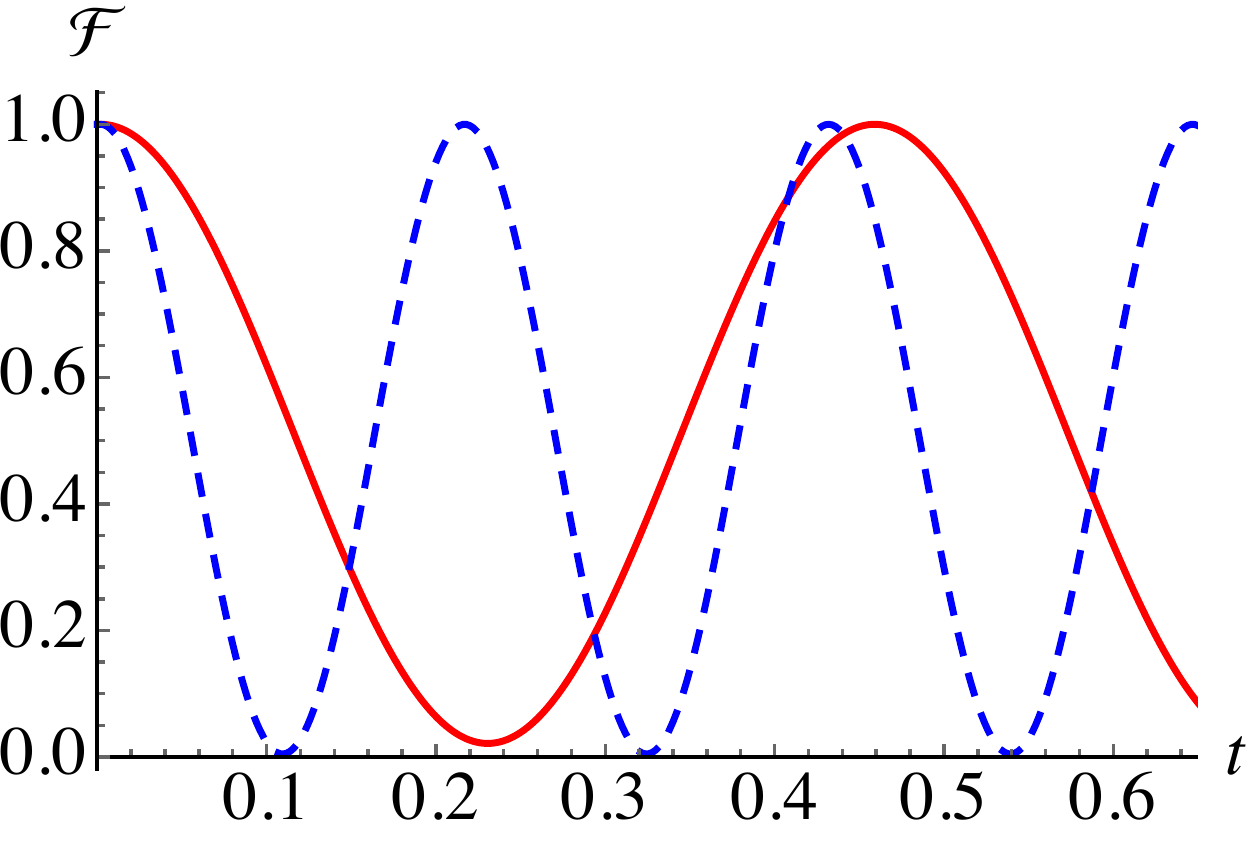}\\
\caption{Survival probability, $\mathcal{F}$, of the impurity+environment state when the LMG bath is initialized with (a) $\lambda\!\!=\!\!0.9$ and (b) $\lambda\!\!=\!\!1.1$ for different values of total number of environmental spins, $N\!\!=\!\!200$ (solid, red) and $N\!\!=\!\!1000$ (dashed, blue).}
\label{fig2}
\end{figure}
We first examine the behavior of $\mathcal{F}$ for the whole system when the interaction, $\gamma$, is suddenly switched on at $t\!\!=\!\!0$. We initialize both in their respective ground states, i.e. for the impurity this simply means that it is always initialized in $\ket{\psi_s}\!=\!\ket{0}_s$, while the ground state of the LMG bath will be dependent on the value of $\lambda$ chosen. For $\lambda\!<\!1$ the field dominates and the spins tend to all align, while for $\lambda\!>\!1$ the ground state is in the critical phase~\cite{HaiTaoPRA2007}.

Quenching on the interaction, $\gamma\!=\!\lambda\sqrt{N}$ drives the system out of equilibrium. In Fig.~\ref{fig2} we examine the survival probability for moderate, $N\!\!=\!\!200$ [solid], and large, $N\!\!=\!\!1000$ [dashed], sized environments for $\lambda\!\!=\!\!0.9$ and $\lambda\!\!=\!\!1.1$, (a) and (b) respectively which are representative values for their phases~\cite{supplementarymaterial}. Clearly for $\lambda\!\!=\!\!0.9$, $\mathcal{F}$ never reaches zero and, furthermore, its behavior is unaffected by  the size of the environment. Therefore, when the LMG model is in this phase we never witness the OC. In contrast, we clearly see that for an environment initialized with $\lambda\!>\!1$ the overall system periodically evolves to almost orthogonal states for moderate sized environments. As we increase the environmental size the minimum value of $\mathcal{F}(t)\!\!\to\!\!0$. Thus, for increasing $N$ the evolved state approaches a fully orthogonal state and the time to reach this state is strongly dependent on the size of the bath, as clearly evidenced in Fig.~\ref{fig2}(b). These features combined indicate that for $\lambda\!>\!1$ the system displays the OC. 

 \begin{figure}[t]
 {\bf (a)} \hskip0.45\columnwidth {\bf (b)}\\
\includegraphics[width=0.49\columnwidth]{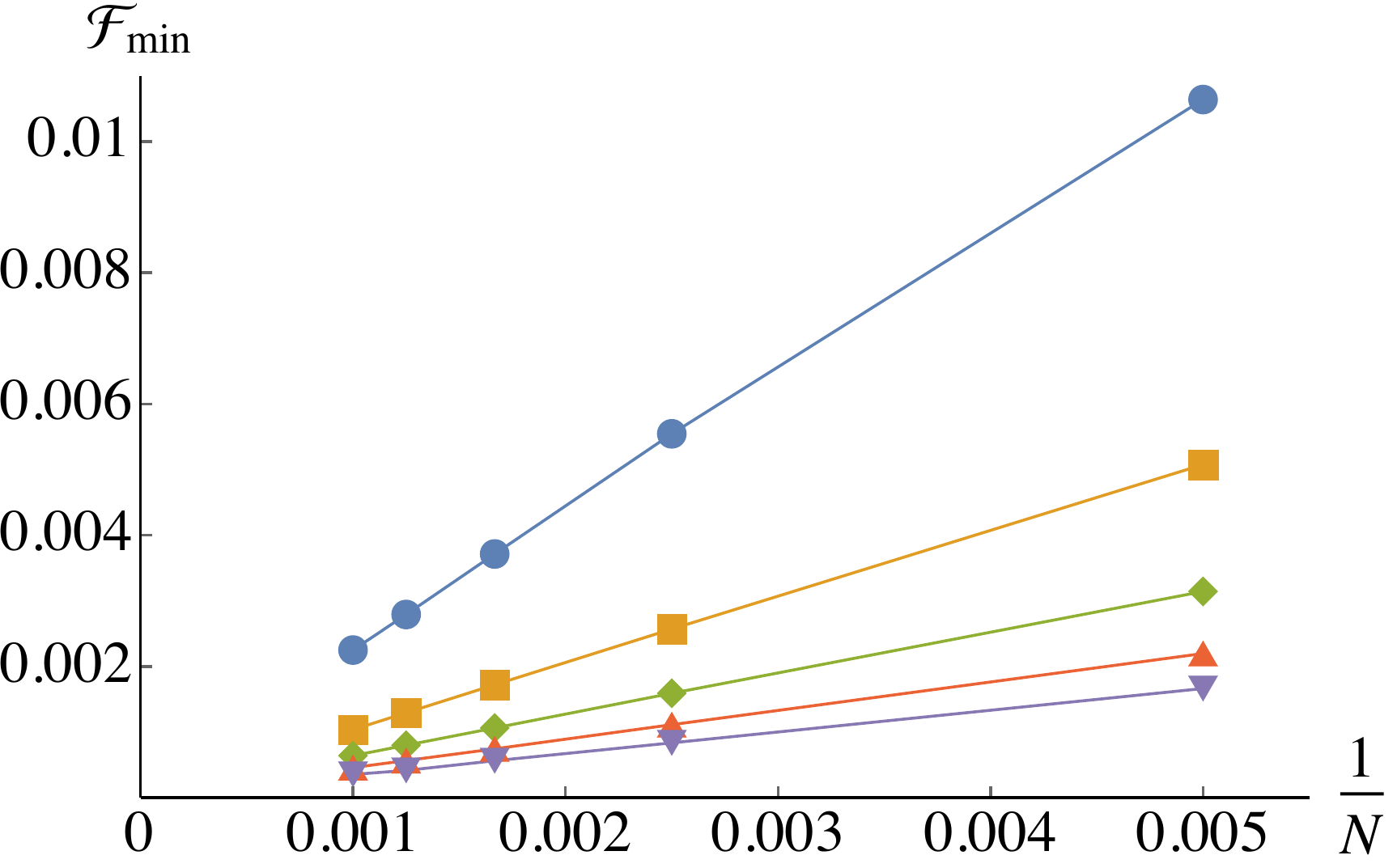}\includegraphics[width=0.49\columnwidth]{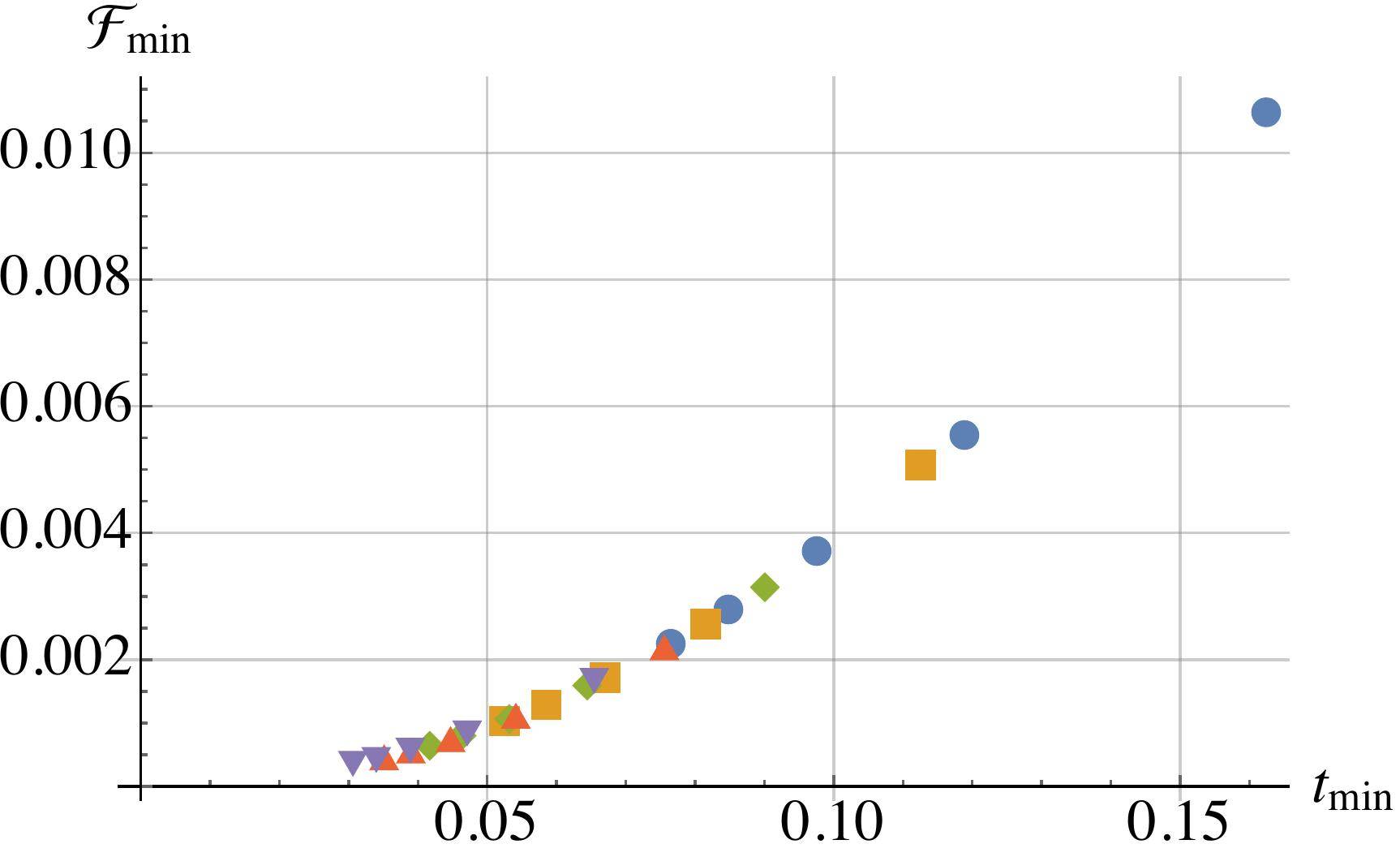}
\caption{(a) $\mathcal{F}_\text{min}$ as a function of $N$. Each line, from top- to bottom-most, corresponds to an increasing value of $\lambda\in(1.2,2.0)$ in steps of 0.2. (b) Minimum value of $\mathcal{F}$ versus corresponding required evolution time, $t_\text{min}$, for different values of $\lambda\!\!=\!\!1.2$ (circles), 1.4 (squares), 1.6 (diamonds), 1.8 (up-triangles), 2.0 (down-triangles). Each consecutive data point approaching the origin corresponds to an increasing value of $N\!\in\!(200,1000)$ in steps of $200$.}
\label{fig3}
\end{figure}

In Fig.~\ref{fig3}(a) we examine the minimum value of the survival probability, $\mathcal{F}_\text{min}$, as a function of inverse environment size, $1/N$ for $\lambda\!>\!1$. Each curve from top to bottom corresponds to an increasingly large value of $\lambda\!\!\in\!\!(1.2,2.0)$. We find a simple linear relationship and it is clear that as $N\!\!\to\!\!\infty$, $\mathcal{F}\!\!\to\!\!0$ and thus we are witnessing the OC. In contrast, when the spin-bath is initialized in the aligned phase the minimal value of the survival probability is insensitive to the bath size, cf. Fig.~\ref{fig2}(a). Fig.~\ref{fig3}(b) shows the relationship between $\mathcal{F}_\text{min}$ and the corresponding time when this minimum occurs, $t_\text{min}$. We clearly see that both $\mathcal{F}_\text{min}$ and $t_\text{min}\!\!\to\!\!0$ as $N\!\!\to\!\!\infty$. Thus, Fig.~\ref{fig3} indicates that when the OC manifests it corresponds to a vanishing orthogonality time as the size of the bath is increased, while if the composite system does not reach orthogonality we find its properties are largely independent of $N$. 

We now would like to connect the above features with the QSL time, Eq.~\eqref{eq:tQSL} and, in particular. shed light onto why despite being a many-body system we do not witness the OC for $\lambda\!<\!1$. In general, the energy spectrum of the LMG model is characterized by a cascade of energy level crossings~\cite{supplementarymaterial} and we find $\Delta H$ reads
\begin{equation}
\label{varF}
\Delta H = \sqrt{ \frac{4(1+j)(N-j)\gamma^2 }{N^2}},
\end{equation}
where $0\!\!\leq\!\! j\!\!\leq\!\! N$ indicates how many energy level crossings have occurred. We find a starkly different behavior depending on which phase the LMG spin bath is initialized in. For $0\!\!<\!\!\lambda\!\!<\!\!1$ no energy level crossings occur~\cite{supplementarymaterial}, and we have $j\!\!=\!\!0$ and $\Delta H \!= \!2 \gamma/ \sqrt{N} $. Therefore, since $\gamma=\lambda \sqrt{N}$ it is clear that the variance is independent of the size of the bath in this phase. The fact that we do not see the OC emerging then naturally follows, since regardless of how large the total system is, the QSL time, Eq.~\eqref{eq:tQSL}, is always the same. We find a very different picture emerging for $\lambda\!\!>\!\!1$ where the energy changes as a function of $\lambda$ due to a cascade of crossings~\cite{supplementarymaterial}. In particular, as $N$ is increased the number of energy level crossings occurring becomes increasingly dense. Thus, we find Eq.~\eqref{varF} scales extensively as the environmental size grows. Correspondingly, we have that $\tau_\text{QSL}\!\!\to\!\!0$ as $N$ grows and hence the orthogonality catastrophe follows as a consequence of the vanishing QSL time.

\paragraph{Concluding remarks.}
In the present analysis we achieved several important results. First and foremost, we related the dynamical occurrence of the orthogonality catastrophe with the quantum speed limit. From a remarkably simple relation, we concluded that any quantum many-body system whose energy variance scales like $N^\alpha$ exhibits the exponential sensitivity to local perturbations. This insight was demonstrated and validated for the trapped Fermi gas, which closely resembles the situation originally studied by Anderson \cite{Anderson}. As a second example, we analyzed the isotropic LMG-model interacting with a single qubit impurity, showing that emergence of the orthogonality catastrophe is dependent on the phase the environment is initialized in. Finally, we also proposed a new QSL that relates the work necessarily performed by the local perturbation for the orthogonality catastrophe to appear. In particular, the last two results may justify further study and encourage the development of a comprehensive thermodynamic framework for quantum many-body systems.

\acknowledgements
T.F. acknowledges support under JSPS KAKENHI-18K13507. S.D. acknowledges support from the U.S. NSF under Grant No. CHE-1648973. This research was supported by grant number FQXi-RFP-1808 from the Foundational Questions Institute and Fetzer Franklin Fund, a donor advised fund of Silicon Valley Community Foundation (S.D.). T.F. and T.B. are supported by the Okinawa Institute of Science and Technology Graduate University. S.C. gratefully acknowledges the Science Foundation Ireland Starting Investigator Research Grant ``SpeedDemon" (No. 18/SIRG/5508) for financial support.

\bibliography{QSL_OC}

\clearpage
\appendix
\section{SUPPLEMENTARY MATERIAL:\\Orthogonality catastrophe as a consequence of the quantum speed limit}
\section{Survival probability behavior as a function of $\lambda$}
As commented in the main text, the behavior of the survival probability shows markedly different characteristics depending on which phase the LMG bath is initialized in. In Fig.~\ref{FigSuppmat}(a) we consider a large environment, $N\!=\!1000$, and sweep through a range of values for $\lambda$. The qualitative and quantitative difference between the two phases is evident. For $\lambda\!<\!1$ we see no evidence of the orthogonality catastrophe, while conversely, as soon as the system is in the critical phase the seemingly small perturbation introduced by quenching on the interaction with the impurity is sufficient to drive the total state to orthogonality. 

\section{Energy spectrum of the LMG model}
\label{LMGspectrum}
In Fig.~\ref{FigSuppmat}(b) we show the energy spectrum for the isotropic LMG model, which is characterized by a cascade of energy level crossings in the critical phase, $\lambda\!>\!1$. We see that in the aligned phase the ground state is a single energy level, and since we are writing eigenstates of the LMG model in terms of the collective spin operator $S_z$, this eigenstate then corresponds to the $\ket{-\tfrac{N}{2}}$ state. When we enter the critical phase we see the energy level crossings. The value of $\lambda$ at which these crossings occur is dependent on the size of the system as evident by comparing the spectra for $N\!=\!10$ and $100$. This cascade has a significant effect on the resulting variance of the energy as discussed in the main text and shown in the inset.

\section{Quantum speed limit and work}
\label{sec:fogartyQSL}
\begin{figure}[b]
{\bf (a)} \hskip0.5\columnwidth {\bf (b)}
\includegraphics[width=0.45\columnwidth]{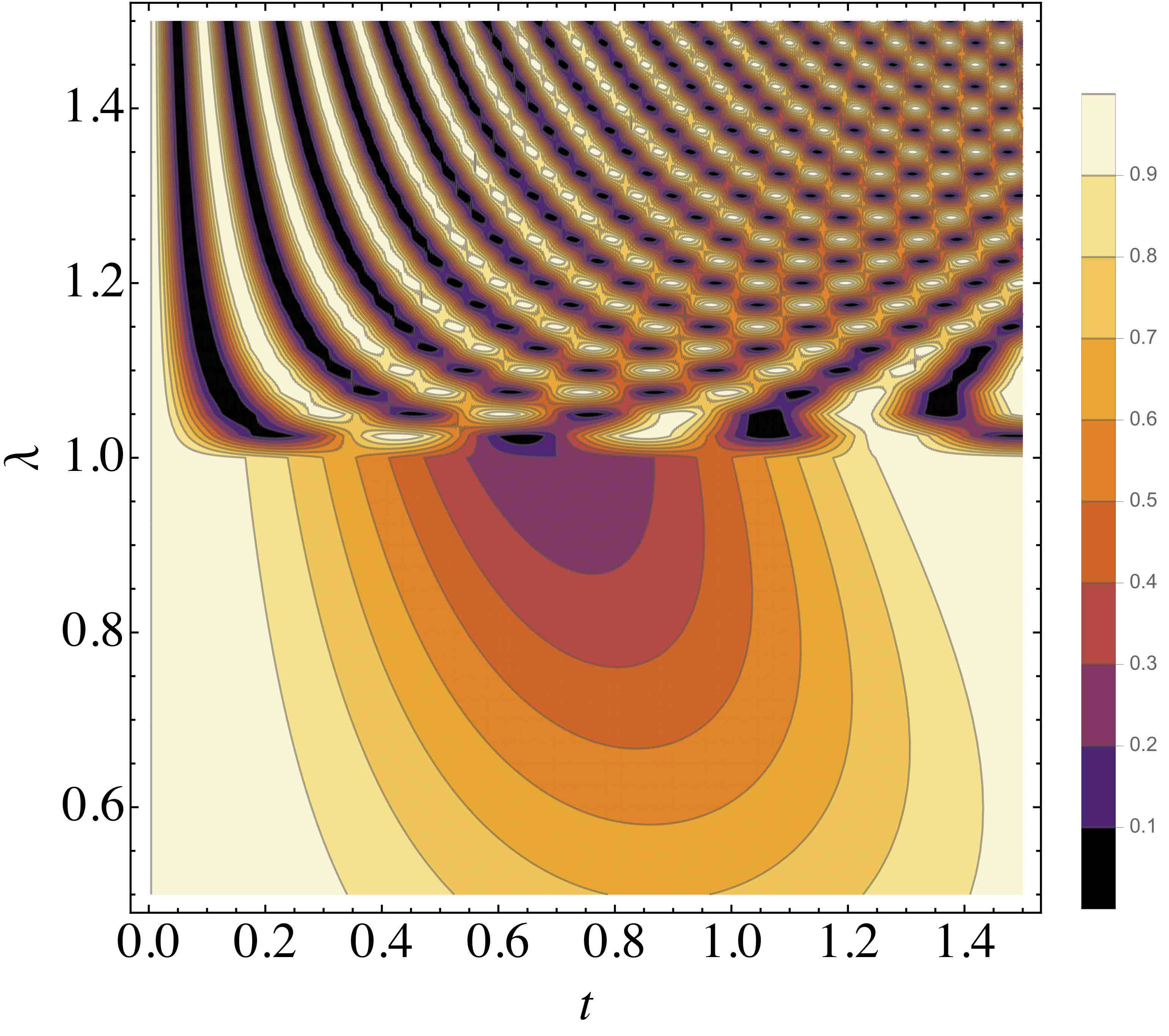}~~\includegraphics[width=0.6\columnwidth]{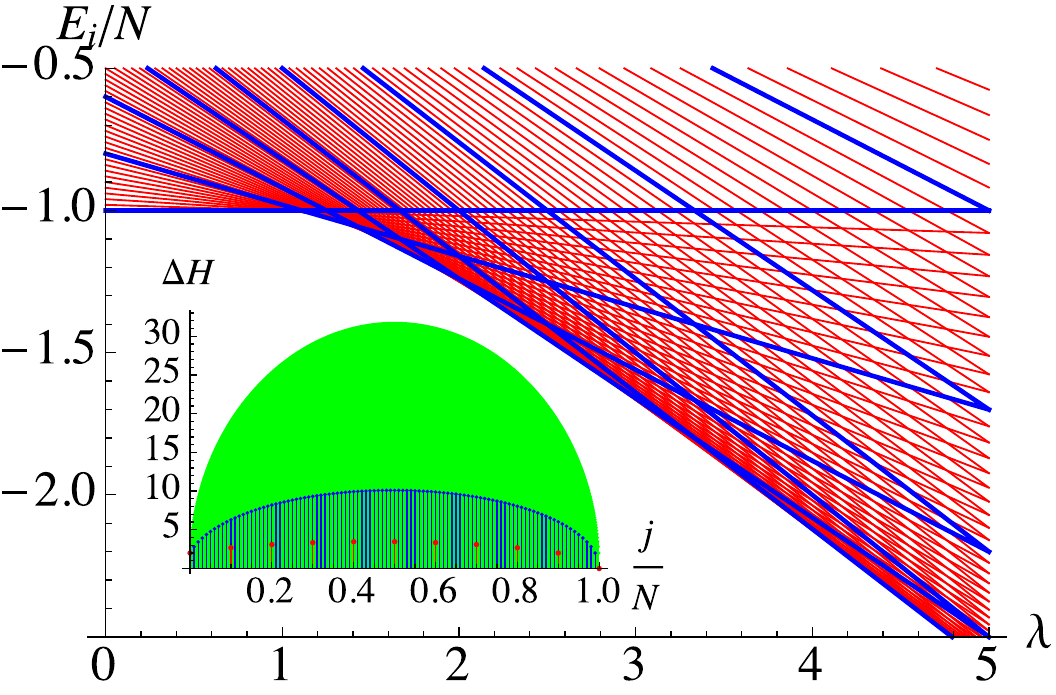}\\
{\bf (c)} \\
\includegraphics[width=0.5\columnwidth]{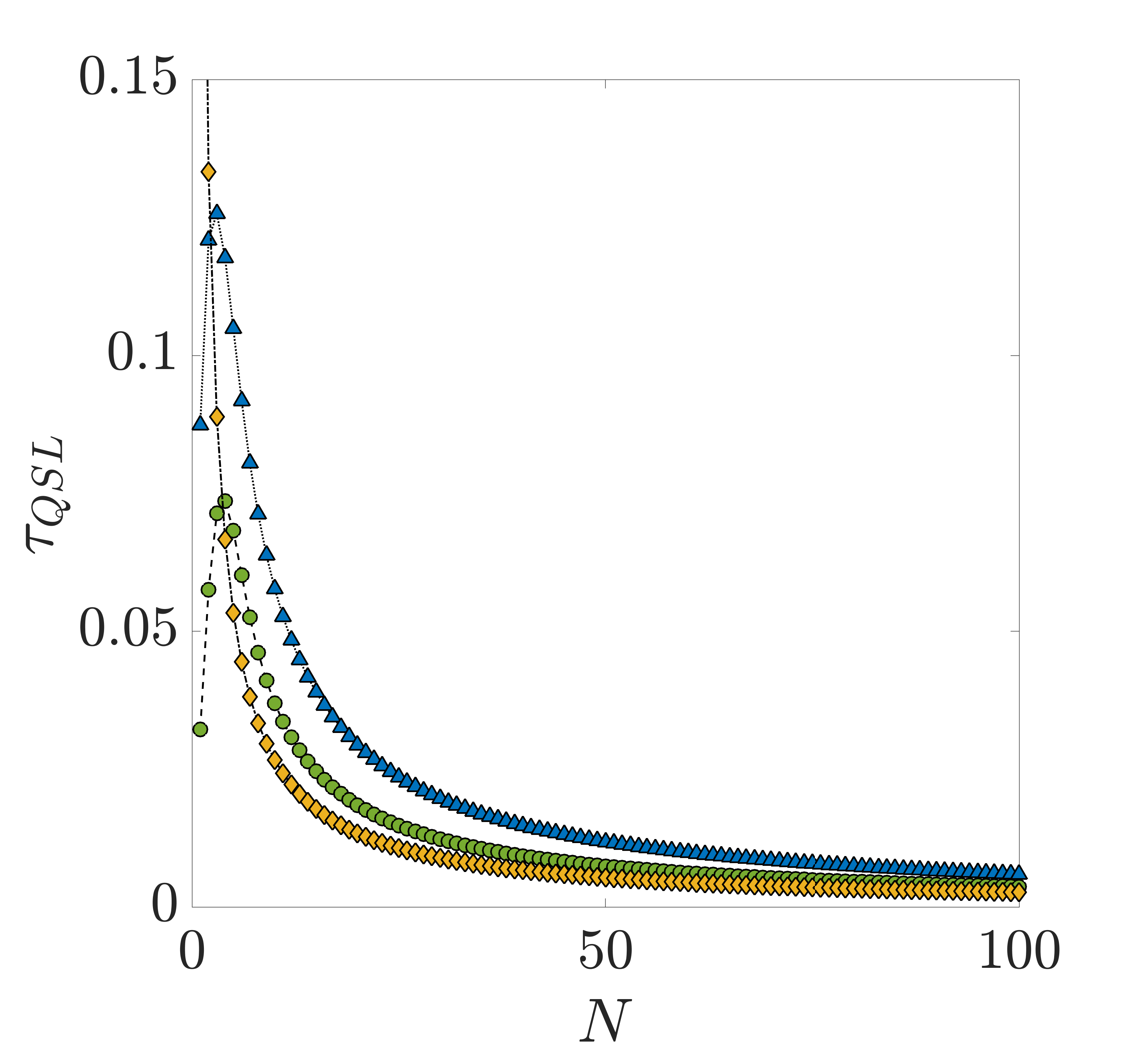}
\caption{(a) $\mathcal{F}$ as a function of time, $t$, and environmental coupling, $\lambda$, for a large environment, $N\!\!=\!\!1000$. (b) Energy spectra for the isotropic LMG model for $N\!=\!10$ (thick, blue) and $N\!=\!100$ (thin, red). {\it Inset:} Behavior of $\Delta H$, Eq.~\eqref{varF} for $N\!=\!10$ (red), $100$ (blue), and 1000 (green). (c) QSL time for the fermionic trap quench of $\omega_2\!=\!4$ with the Mandelstam-Tamm (blue triangles), Margolus-Levitin (green circles), and the average work bound (yellow diamonds).}
\label{FigSuppmat}
\end{figure}

Here we provide some additional details for deriving Eq.~\eqref{fogartyQSL}. Consider again the Bures angle $\mc{L}$ for which we have that 
\begin{equation}
\partial_t\mathcal{L}(\ket{\psi_0},\ket{\psi_\tau}) \leq \vert \partial_t \mathcal{L}(\ket{\psi_0},\ket{\psi_\tau})   \vert = \left\vert \frac{ \partial_t \bra{\psi_0}\psi_t\rangle }{\sqrt{1-\bra{\psi_0}\psi_t\rangle^2 }}  \right\vert
\end{equation}
and therefore
\begin{equation}
 \dot{\mathcal{L}} \sin\mathcal{L} \leq \vert  \partial_t  \bra{\psi_0}\psi_t\rangle \vert.
\end{equation}
Taking the spectral decomposition of the initial (final) total Hamiltonian, $H_{i(f)}=\sum_j \tfrac{E^{i(f)}_j}{\hbar} \ket{\phi^{i(f)}_j}\!\bra{\phi^{i(f)}_j}$. For a sudden quench the evolved state is simply given by
\begin{equation}
\ket{\psi_t} = \sum_j e^{-i E^{f}_j t/\hbar} \ket{\phi^{f}_j}\!\bra{\phi^{f}_j} \psi_0 \Big>.
\end{equation}
Assuming the system is initially in the ground state, the survival probability is essentially the absolute value squared of the characteristic function of the work distribution~\cite{Silva2008,Talkner2016}
\begin{equation}
\chi(t) =  \sum_j e^{i (E^{f}_j - E^i_0)t/\hbar}\, \Big\vert \bra{\phi^{f}_j} {\psi_0}\Big>\,\Big\vert^2,
\end{equation}
and we obtain
 \begin{equation}
\dot{\mathcal{L}} \sin\mathcal{L} \leq \vert  \partial_t  \chi(t) \vert = \left\vert \sum_j \frac{(E^{f}_j-E^{i}_0)}{\hbar} e^{i (E^{f}_j - E^i_0)t/\hbar} \left\vert \bra{\phi^{f}_j}\psi_0\Big> \right\vert^2  \right\vert.
\end{equation}
Integrating and simplifying we find
\begin{equation}
\label{fogartyQSL2}
\tau_\mrm{W} = \frac{\hbar\left(1 - \vert\chi(t)\vert \right)}{\vert \langle W \rangle \vert}.
\end{equation}
where $\langle W \rangle=\partial_t \chi(t) \vert_{t=0}$ is the average work done in performing the quench. Indeed for the simple case of the trap quench with the non-interacting Fermi gas, the average work is given as $\langle W \rangle=\frac{N}{4}\left[\eta^2-1\right]$ therefore exhibiting similar scaling to $\tau_\mrm{QSL}$, see Fig.\ref{FigSuppmat}(c).

\end{document}